\documentclass[11pt,twocolumn]{article}
\usepackage[a4paper,margin=1in]{geometry}
\usepackage[T1]{fontenc}
\usepackage[utf8]{inputenc}
\usepackage{lmodern}
\usepackage{microtype}
\usepackage{xcolor}
\usepackage{booktabs}
\usepackage{array}
\usepackage{enumitem}
\usepackage{graphicx}
\usepackage{amsmath}
\usepackage{amssymb}
\usepackage{seqsplit}
\usepackage{xspace}
\usepackage{authblk}
\usepackage{caption}
\usepackage[numbers,sort&compress]{natbib}
\usepackage{hyperref}
\usepackage{upquote}

\hypersetup{
  colorlinks=true,
  linkcolor=blue!45!black,
  citecolor=blue!45!black,
  urlcolor=blue!55!black,
  pdftitle={Aray: Deterministic-First Synthesis of Benign Artifacts for YARA Validation},
  pdfauthor={Emanuel Valente, Lourenco A. P. Junior, Leonardo Chahud, Julio Cezar Estrella, Marcus Botacin},
  pdfsubject={Deterministic synthesis of positive scanner fixtures for YARA validation},
  pdfkeywords={YARA, binary synthesis, ELF, PE, malware analysis, detection engineering}
}

\usepackage{listings}
\definecolor{listingblue}{HTML}{1F4E79}
\definecolor{listinggreen}{HTML}{386641}
\definecolor{listingbrown}{HTML}{7A3E00}
\definecolor{listinggray}{HTML}{5F6368}
\definecolor{listingbg}{HTML}{F7F8FA}

\lstdefinelanguage{YARA}{
  sensitive=true,
  morekeywords={rule,private,global,strings,condition,meta,import,include,
    and,or,not,all,any,of,them,at,in,filesize,for,uint16,uint32,
    uint16be,uint32be,ascii,wide,nocase,fullword},
  morecomment=[l]{//},
  morecomment=[s]{/*}{*/},
  morestring=[b]"
}

\lstdefinelanguage{GNUAssembler}{
  sensitive=true,
  morekeywords={.section,.byte,.globl,.text,.data,.bss,leaq,movq,xorq,
    syscall},
  morecomment=[l]{\#}
}

\lstdefinelanguage{LinkerScript}{
  sensitive=true,
  morekeywords={PHDRS,SECTIONS,PT_LOAD,FILEHDR,SIZEOF_HEADERS},
  morecomment=[s]{/*}{*/}
}

\lstdefinestyle{aray}{
  basicstyle=\ttfamily\footnotesize,
  keywordstyle=\color{listingblue}\bfseries,
  commentstyle=\color{listinggreen}\itshape,
  stringstyle=\color{listingbrown},
  numberstyle=\scriptsize\color{listinggray},
  columns=fullflexible,
  breaklines=true,
  breakatwhitespace=false,
  keepspaces=true,
  showstringspaces=false,
  upquote=true,
  tabsize=2,
  frame=tb,
  framerule=0.45pt,
  rulecolor=\color{black!35},
  backgroundcolor=\color{listingbg},
  framesep=5pt,
  xleftmargin=1em,
  xrightmargin=0.5em,
  framexleftmargin=0.5em,
  aboveskip=0.9em,
  belowskip=0.6em,
  captionpos=b
}
\lstdefinestyle{araynumbered}{style=aray,numbers=left,numbersep=8pt}
\newcommand{\tool}{\textsc{Aray}\xspace}
\newcommand{\code}[1]{\texttt{#1}}

\makeatletter
\renewcommand\section{\@startsection{section}{1}{\z@}%
  {-2.2ex \@plus -.8ex \@minus -.2ex}%
  {0.9ex \@plus .2ex}%
  {\normalfont\large\bfseries}}
\renewcommand\subsection{\@startsection{subsection}{2}{\z@}%
  {-1.8ex \@plus -.6ex \@minus -.2ex}%
  {0.6ex \@plus .2ex}%
  {\normalfont\normalsize\bfseries}}
\renewcommand\subsubsection{\@startsection{subsubsection}{3}{\z@}%
  {-1.5ex \@plus -.5ex \@minus -.2ex}%
  {0.5ex \@plus .2ex}%
  {\normalfont\normalsize\itshape}}
\makeatother

\renewenvironment{abstract}{%
  \par\small\bfseries
  \renewcommand{\tool}{\textup{Aray}\xspace}%
  \noindent\textit{\abstractname}\textemdash\ignorespaces
}{%
  \par\normalfont\normalsize\vspace{0.6em}
}

\title{%
  \textbf{Aray: Deterministic-First Synthesis of Benign Artifacts for YARA Validation}%
  \thanks{The tool and this work were presented at
    \href{https://blackhat.com/us-26/arsenal/schedule/\#aray-benign-binary-synthesis-for-signature-validation-without-the-malware-52919}{Black Hat USA 2026 Arsenal}.}
}
\author[1,2]{Emanuel C. A. Valente}
\author[3]{Lourenço A. P. Júnior}
\author[3]{Leonardo Gonçalves Chahud}
\author[2]{Júlio Cezar Estrella}
\author[4]{Marcus Botacin}
\affil[1]{iFood}
\affil[2]{University of São Paulo (USP)}
\affil[3]{Aeronautics Institute of Technology (ITA)}
\affil[4]{Texas A\&M University (TAMU)}
\date{%
  \vspace{-0.8em}\footnotesize
  \texttt{emanuel.valente@usp.br}\quad
  \texttt{ljr@ita.br}\quad
  \texttt{chahud@ita.br}\\[-0.1em]
  \texttt{jcezar@icmc.usp.br}\quad
  \texttt{botacin@tamu.edu}
}

\begin{document}
\maketitle
\vspace{-1.2em}

\begin{abstract}
A YARA rule is easy to distribute, but the malware sample used to demonstrate a
positive match is not. This complicates safe storage, continuous integration,
disaster-recovery exercises, and reproducible scanner validation. Constructing a
replacement fixture requires more than embedding literals: YARA conditions can
combine alternatives, counts, offsets, integer reads, and executable-container
constraints, while the resulting file should not reproduce malware behavior.

Our insight is that positive validation is existential: it requires one
file-level member of a rule's match set, not reconstruction of the originating
sample. We present \tool, a deterministic-first YARA interpreter and
positive-fixture synthesizer. Models may propose constructive normalizations or
typed extraction fallbacks, but never backend source or binary structure.
Conventional code canonicalizes and validates normalized rules for syntax, rule
identity, retained declarations and modifiers, supported count expansions, and
fixed witnesses
derived from regex and variable-hex declarations. It then performs extraction,
routing, collision-checked layout, and ELF, PE, or generic serialization; only
residual normalization semantics reach a bounded model judge.

We evaluated \tool over 416 selected public-rule entries. Normalization accepted
182 entries without model assistance and 234 after model normalization.
Constructibility preflight admitted 406 entries, and every admitted fixture
matched its associated upstream original rule. This yields 406/416 (97.6\%)
overall and 406/406 among constructible rules, with ten expected preflight
dispositions and zero \code{yara\_mismatch} or construction failures. An
unreachable endpoint confirmed zero model invocations during realization. The
original-rule oracle therefore provides direct source-rule validation for every
generated fixture; proving implication for all possible files is a separate,
stronger objective. Two anchored-regex failures discovered with this corpus were
repaired before the final run; the reported figures are therefore post-fix
systems results, not a held-out estimate of normalization generalization.
\end{abstract}

\section{Introduction}

YARA provides a compact language for describing byte patterns, strings,
offsets, counts, and structural properties of files. It is widely used for
malware intelligence and detection engineering, and prior work has addressed
scalable rule deployment and automatic rule generation
\cite{yaradocs,brengel2021yarix,raff2020autoyara,li2023packgenome,mansour2025neuroyara,coscia2025apiary}.
The positive validation of a rule, however, commonly remains coupled to the
availability of a representative malware sample.

That coupling has operational costs. A live-malware corpus is not merely a
storage allocation: it requires provenance and retention policy, access control,
controlled transfer and replication, and an isolated analysis environment whose
network policy, credentials, snapshots, and teardown procedures must themselves
be maintained. Samples may consequently be unavailable in continuous-integration
or air-gapped environments, and sharing a rule is generally easier than sharing
the specimen that demonstrates its behavior. A rule can therefore be
syntactically valid and operationally deployed without a portable positive test
fixture.

The problem becomes sharper during disaster-recovery exercises. Recovery
guidance calls for realistic scenarios that test and iteratively improve the
recovery plan, rather than merely establishing that backup objects exist
\cite{bartock2016recovery}. A useful rehearsal of a detection service must show
that a restored or alternate environment can ingest an object, emit the expected
event, load the intended rule set, quarantine the object, and deliver an alert.
Replicating live malware into recovery accounts or regions expands its custody
boundary; excluding it leaves the positive path untested. Maintaining a second
malware-capable execution enclave solely for such rehearsals further increases
cost and attack surface.

This gap also appears in industrial object-storage pipelines. An organization
may deploy YARA scanning behind uploads to Amazon S3 or an S3-compatible store,
where an object-created event triggers scanning, quarantine, and alerting. A
generated fixture can be uploaded to a dedicated test bucket in either the
primary or recovered environment to exercise that same ingestion path---event
delivery, scanner configuration, rule deployment, quarantine policy, and alert
transport---without placing live malware in the bucket. \tool does not implement
storage integration or disaster recovery; it supplies the portable positive
scanner fixture needed to test those systems.

LLM-assisted cyber testing also raises containment concerns when models receive
tools and long execution horizons. The July 2026 OpenAI--Hugging Face incident
shows that nominal isolation and restricted egress can fail when an agent chains
vulnerabilities across trusted services
\cite{openai2026hfincident,huggingface2026timeline}; this motivates withholding
shell, direct network access, and binary-construction authority from the model
rather than relying on sandboxing alone.

This work studies a narrower alternative to malware emulation: synthesize a
file whose observable bytes satisfy a YARA rule, without recreating the
malware's behavior. In this formulation, the artifact is a witness to a set of
file-layout constraints. It may contain family names, byte signatures, magic
values, or encoding-specific strings because the rule requires them, but it
does not intentionally implement persistence, command and control, exploitation,
or payload logic. The term \emph{benign} in the title denotes that design intent;
it is not a general proof that arbitrary matched content is harmless.

We present \tool, a deterministic-first YARA interpreter and positive-fixture
synthesizer. Its central design decision is to place conventional code, not a
generative model, in authority over backend syntax, container structure, and
serialization. A model may select or propose payload literals when simplifying
unsupported syntax, but it never emits backend source code, container metadata,
or a complete binary, and Aray grants it no shell, storage credential, or network
tool beyond the configured inference request. This separation makes the system
inspectable and permits a zero-inference execution path for already normalized
rules, although the current graph still initializes model clients. It narrows
model authority inside Aray; it is not a substitute for isolation of the wider
test environment.

The contributions of this work are:
\begin{itemize}[leftmargin=1.4em]
  \item a deterministic methodology for reducing a YARA source file to one
  standalone target rule, assessing constructibility, and extracting typed
  string and integer witnesses;
  \item a deterministic validation layer for normalized rules, covering YARA
  syntax, selected-rule identity, retained declarations and modifiers, supported
  count expansions, contextual regex witnesses, and hex witnesses, with model
  judging reserved for residual semantics;
  \item executable ELF and PE synthesis techniques for exact file offsets,
  including a GNU linker-script identity for ELF and a two-pass low-alignment
  PE construction with post-build verification;
  \item direct scanner-oriented writers, which create ELF, PE, or generic
  artifacts for scanning without requiring execution or a compiler toolchain;
  and
  \item a staged end-to-end evaluation over 416 public-rule entries, using a
  derivative-independent upstream-original-rule YARA oracle, a frozen handoff,
  and an unreachable model endpoint to isolate deterministic realization.
\end{itemize}

\begin{figure}[t]
  \centering
  \includegraphics[width=0.96\linewidth]{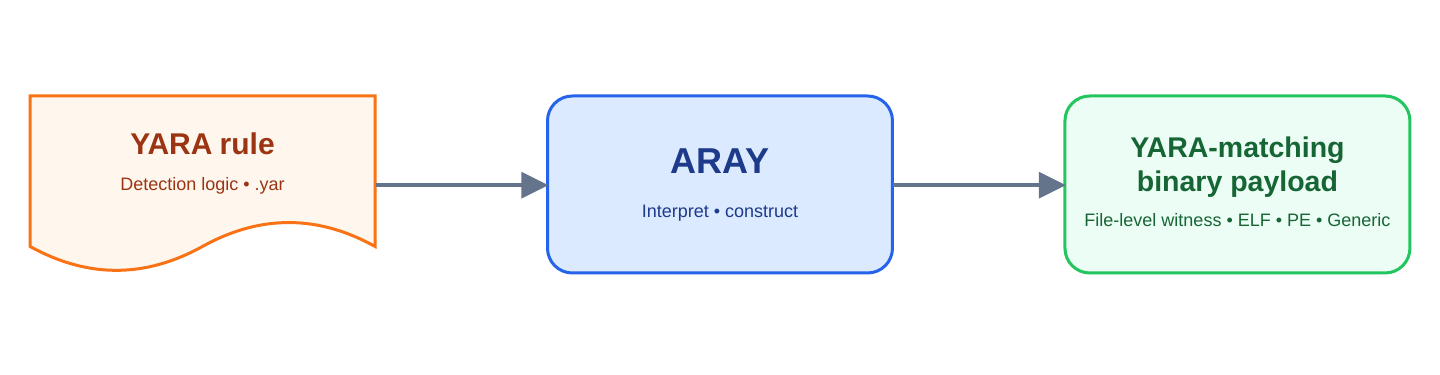}
  \caption{Conceptual view of \tool: a YARA rule is realized as a YARA-matching
  binary payload. The output is matching-equivalent to a positive sample only
  under the selected file-level predicate; it does not reproduce behavior. The
  evaluation checks this predicate with the selected upstream original rule.}
  \label{fig:overview}
\end{figure}

\section{Problem Definition and Scope}

Let $r$ be a selected YARA rule and $b$ a byte sequence. The synthesis objective
is
\begin{equation}
  \label{eq:synthesis-objective}
  r(b)=\mathsf{match}.
\end{equation}
Equivalently, let
\begin{equation}
  \mathcal{M}_r=\{b\mid r(b)=\mathsf{match}\}
\end{equation}
be the rule's match set. \tool seeks one witness $b\in\mathcal{M}_r$; it does
not enumerate that set, establish a one-to-one correspondence with source
samples, or identify a unique canonical member. Fixed strings may admit many
layouts, and regular expressions or variable hex patterns may admit infinitely
many byte sequences. The implemented policies choose one constructible candidate
using stable order, minimum-length choices, and backend feasibility, not a
globally optimal representative.

The generated $b$ is a positive scanner fixture, not a reconstruction of the
program from which the rule was derived. For model-normalized rules, \tool aims
to produce a sufficient subset $r'$ such that
\begin{equation}
  \label{eq:constructive-specialization}
  \forall x,\quad r'(x)=\mathsf{match}\Rightarrow r(x)=\mathsf{match},
\end{equation}
while permitting $r'$ to match fewer files than $r$. General implication over
the full YARA language is not proven. The system combines deterministic proofs
for a narrow fragment, deterministic witness checks with the YARA engine, and,
when necessary, a bounded model judge. Separately, the evaluation scans the final
artifact with the upstream original rule. This is an independent test of the
concrete witness $b$, not a proof that Equation~\ref{eq:constructive-specialization}
holds for every possible byte sequence.

The supported deterministic fragment has three main parts. Its string subset
covers fixed text or hex data and selected encoding, case, and word-boundary
modifiers. Its placement subset covers direct references, simple set
quantifiers, literal offsets or ranges, and lower bounds on occurrences. Its
integer subset covers 16- and 32-bit unsigned reads in either byte order,
literal comparisons, and selected nested expressions based on a little-endian
\code{uint32} pointer. The parser records \code{int16} metadata, but signed
witness selection is incomplete and is not part of the claimed sound fragment.
The backends also recognize conjunctive textual \code{filesize} bounds. Here,
\emph{sound fragment} means that the stated construction strategy supplies
sufficient evidence for an accepted form, subject to the qualifications in
Table~\ref{tab:extract-fragment}.

The current system is not a complete YARA solver. It does not interpret imported
module semantics or solve general arithmetic and Boolean conditions. Arbitrary
loops, percentage counts, and prescribed whole-file cryptographic hashes are
also outside its constructible subset. Recognized complex forms can request
normalization or extraction fallback, but the lexical capability detector is
not complete. An unrecognized form may therefore be over-constrained or proceed
with incomplete evidence. Independently, cross-rule processing is deliberately
one-target: the first non-private rule and its reachable dependencies are
selected, while unrelated rules are excluded.

\section{Architecture and Trust Boundary}

\begin{figure*}[t]
  \centering
  \includegraphics[width=0.95\textwidth]{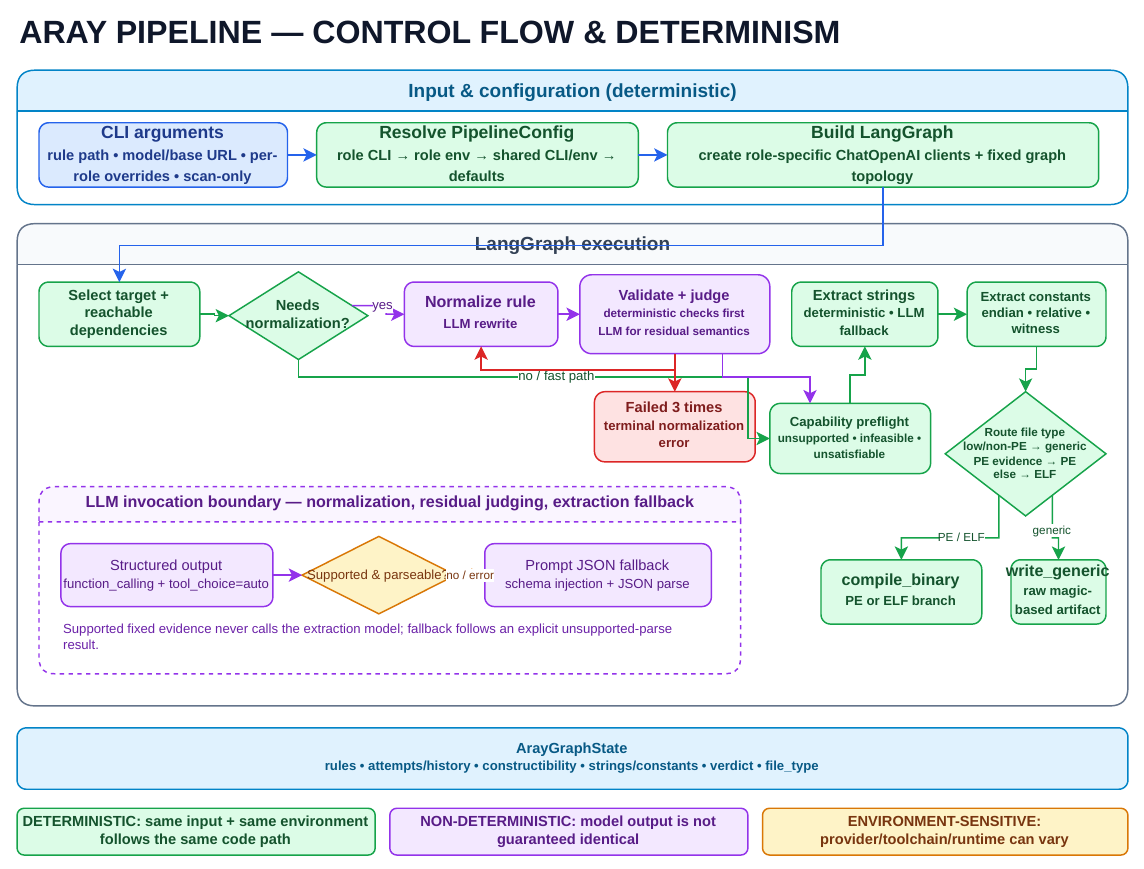}
  \caption{Control flow and determinism boundary. Normalized candidates first
  undergo deterministic syntax and supported-semantic validation, including
  contextual regex and hex witness checks. Supported syntax then follows
  capability preflight and deterministic extraction. Model calls are restricted
  to proposal generation, residual semantic judging, and unsupported extraction
  syntax. Green denotes conventional code, purple denotes a possible model
  invocation, and orange denotes environment-sensitive behavior.}
  \label{fig:pipeline}
\end{figure*}

Figure~\ref{fig:pipeline} separates deterministic code, model-assisted stages,
and environment-sensitive toolchains. LangGraph provides control flow and
bounded retries; it does not construct the artifact. The trust boundary is the
typed representation between interpretation and construction.

The pipeline state records the selected and normalized source together with its
constructibility and extraction provenance. It also carries typed evidence,
normalization history, and the target format. A successful deterministic parse
is authoritative: extraction fallback is not asked to reinterpret or override
it. The binary backend receives only typed entries and the rule text needed for
routing and filesize constraints.

\subsection{Source Reduction}

The front end uses a structural lexer that treats quoted literals, regular
expressions, and comments as opaque tokens. It identifies top-level imports,
rules, and section spans without allowing keywords inside data to influence the
parse. The first non-private rule is selected as the target.

Bare references to helper rules are resolved recursively from the same file or
sibling \code{.yar} files. Helper conditions are inserted parenthesically, and
their string declarations are merged into the target under stable namespaces.
For example, \code{\$a} from rule \code{Helper} becomes
\code{\$\_\_Helper\_a}; references through \code{\$}, \code{\#}, \code{@},
and \code{!} are renamed consistently. Anonymous target strings receive stable
names of the form \code{\$\_\_aray\_anon\_N} before any possible model call.
Cycles, ambiguous sibling declarations, and unresolved dependencies terminate
processing explicitly.

Rule-set quantifiers are reduced to one sufficient deterministic witness:
\code{all} becomes the conjunction of every matching helper, while \code{any},
numeric, and percentage quantifiers select the first sufficient declarations in
stable source order. This reduction preserves a constructive witness but is not
an equivalent expansion of every alternative. Global rule conditions are
conjoined with the selected target.

\section{Rule Normalization Strategy}
\label{sec:normalization}

The normalization strategy is a component introduced by this work; it is not a
service provided by the YARA engine. Its purpose is to bridge the gap between a
rule that describes a family of acceptable files and a backend that must realize
one concrete file. Regular expressions and variable hex patterns denote many
possible byte sequences. Disjunction and count thresholds likewise admit
multiple evidence combinations. Requiring every alternative can turn a feasible
rule into an impossible construction, while selecting fragments without regard
to Boolean structure can produce a file that does not satisfy the source rule.

Normalization is an Aray subsystem, not external corpus preprocessing. The
regular pipeline invokes it as a graph node, while the \code{aray-normalize}
batch interface shares source selection, the \code{normalize\_rule}
implementation, and deterministic candidate checks. Its residual-judge admission
is stricter: the batch retains only \code{passed} candidates, whereas the regular
graph currently proceeds on \code{uncertain}. The evaluation uses the stricter
batch policy to expose normalization provenance and then freezes its exact
accepted outputs before artifact realization. Final acceptance is not based only
on those derivatives: the evaluator independently associates and scans the
corresponding upstream original rule.

\subsection{Constructive Specialization}

Rather than claim semantic equivalence, normalization seeks a
\emph{constructive specialization} $r'$ satisfying the direction in
Equation~\ref{eq:constructive-specialization}. The candidate may accept fewer
files than $r$, but every file it accepts should also satisfy $r$. This direction
permits one complete disjunctive branch to stand for an \code{or}, or one valid
combination of declarations to stand for an \code{N of} threshold. For example,
the condition \code{(\$a and \$b) or (\$c at 0x200)} can specialize to the
complete right branch; combining \code{\$a} with \code{\$c} would not preserve
either branch.

This formulation is deliberately different from rewriting a rule into a more
permissive approximation. It reduces choice while retaining sufficient evidence
for a positive witness. General implication over YARA is not decided by the
implemented method and is not claimed. The system instead combines deterministic
proofs for a narrow fragment with candidate validation and bounded residual
judging.

\subsection{Triggering and Candidate Generation}

Normalization is not the default extraction mechanism. A deterministic lexical
classifier requests it for regular expressions, hex wildcards, and hex jumps.
Disjunction and numeric \code{N of} expressions also trigger this path. A rule
without these indicators is compiled with YARA for basic validation and proceeds
directly to constructibility preflight.

When normalization is required, the model receives the selected standalone rule
and is asked for a constructible specialization, not a binary. Boolean
simplification must retain a complete branch under YARA precedence. Threshold
expansion chooses enough explicit declarations to satisfy the count. Pattern
concretization replaces a regular expression or variable hex pattern with bytes
accepted by that original declaration. The prompt also favors evidence that is
compatible with format, offset, and filesize constraints. These choices can be
non-unique, which is why candidate generation remains outside the deterministic
core. Regex anchors are treated as positional assertions: an unanchored
alternative is preferred, and a start-anchored witness is accepted only when an
exact \code{at 0} constraint is preserved. The current validator rejects fixed
replacements that depend on end-of-file placement; it does not derive equivalent
EOF constraints from general \code{filesize} expressions.

\subsection{Canonicalization and Deterministic Validation}

The model output is a proposal. Three layers constrain it before construction:
\begin{enumerate}[leftmargin=1.6em]
  \item \textbf{Canonicalization.} Conventional code restores retained fixed
  literals, fixed hex declarations, and recognized modifier sets exactly. It
  also repairs safe anonymous aliases and expands one unambiguous exact count.
  For linear hex
  patterns---patterns whose wildcard and jump choices can be resolved without
  alternatives or dependencies---it derives a canonical witness. For example,
  \code{A?} becomes \code{A0}, \code{?B} becomes \code{0B}, and a range
  \code{[n-m]} uses its minimum length $n$.
  \item \textbf{Validation and proof.} The candidate must compile with
  \code{yara-python}, contain exactly one public rule, and retain the selected
  identifier. Deterministic checks reject added declarations, changed modifiers
  or fixed values, invalid anonymous aliases, incorrect count expansions, and
  fixed witnesses derived from regex or hex declarations that do not satisfy the
  supported checks. A
  narrow top-level conjunction and count fragment is proven deterministically
  and bypasses model judging.
  \item \textbf{Residual judge.} Only candidates that pass deterministic checks
  but fall outside the proof fragment are sent to a model judge. A failed verdict
  supplies feedback and retries normalization up to three times. Three
  failures stop construction; an uncertain verdict currently proceeds.
\end{enumerate}

Regex validation is placement-sensitive rather than a match against the
candidate bytes in isolation. For each retained regex, conventional code tests
the permitted ASCII and wide encodings in representative start, end, interior,
word-neighbor, and non-word-neighbor contexts. A top-level exact \code{at}
constraint restricts the tested placement class. At each context offset $k$, an
auxiliary YARA rule searches for a counterexample of the form
\code{candidate at k and not original at k}. Any such match rejects the proposal.
This prevents an end-anchored or doubly anchored regex from becoming an
unconstrained fixed literal, while still accepting a start-anchored witness when
\code{at 0} is preserved. The context family is a deterministic check for the
supported boundary cases, not an exhaustive formal proof over every regex
context.

The resulting normalized rule is preserved with the artifact as the exact
construction predicate used for extraction and realization. This provenance is
important because deterministic validation does not prove general subset
implication. The strategy therefore isolates uncertainty rather than presenting
normalization as a semantics-preserving compiler pass. In the reported
evaluation, acceptance adds a separate layer: the completed artifact must match
the independently selected upstream original rule.

\section{Deterministic Extraction Methodology}
\label{sec:extraction}

The extraction phase converts a normalized rule into two typed collections. A
string entry is represented conceptually as
\begin{equation}
  \begin{aligned}
    s={}&(id,x,e,o,[l,u],c,\\
        &\quad A,W,F,N),
  \end{aligned}
\end{equation}
where $x$ is the witness value, $e$ its ASCII/hex/wide encoding, $o$ an exact
offset, and $[l,u]$ an optional placement interval. The count $c$ controls
multiplicity, while $A,W,F,N$ retain the ASCII, wide, fullword, and nocase
properties. A constant entry is
\begin{equation}
  k=(v,o,w,E,S,D,\delta),
\end{equation}
where $v$ is the concrete value and $o$ the innermost literal offset. The tuple
$(w,E,S)$ describes the outer read width, byte order, and signedness metadata;
$(D,\delta)$ describes nesting and relative displacement. Pydantic validates
field widths and placement ranges. It also prevents simultaneous exact and
ranged placement and rejects incomplete hexadecimal bytes. Deterministic
parsing supplies non-negative exact offsets, but the fallback schema does not
currently enforce that invariant for string offsets.

Table~\ref{tab:extract-fragment} states the operational extraction contract. It
distinguishes a witness strategy from full semantic equivalence: some accepted
forms deliberately construct one sufficient witness, and exact-count absence is
not globally proven.

\begin{table*}[t]
  \small
  \centering
  \begin{tabular}{@{}p{0.24\textwidth}p{0.32\textwidth}p{0.36\textwidth}@{}}
    \toprule
    \textbf{Normalized form} & \textbf{Deterministic action} & \textbf{Qualification} \\
    \midrule
    Fixed literal or fixed hex & Preserve concrete bytes and modifiers & ASCII wide is preferred when both encodings are allowed \\
    Direct \code{\$id} / immediate \code{not \$id} & Include / exclude declaration & Parenthesized or compound negation is outside the sound fragment \\
    \code{all}/\code{any} set & Include all / first stable member & \code{any} is a sufficient, non-unique witness \\
    Literal \code{at} & Duplicate entry for every offset & Every exact string interval must be non-overlapping \\
    Literal \code{in} & Search first free offset in one interval & At most one retained interval per identifier \\
    Simple \code{\#id} comparison & Emit intended occurrence lower bound & Exact equality can be invalidated by incidental matches \\
    Unsigned integer comparison & Select a width-bounded witness & Big-endian serialization is retained \\
    Nested \code{uint32} pointer & Retain base and non-negative displacement & Broader inner widths/orders are not in the sound subset \\
    \bottomrule
  \end{tabular}
  \caption{Deterministic extraction fragment and its witness semantics.}
  \label{tab:extract-fragment}
\end{table*}

\subsection{Constructibility Preflight}

After optional normalization and before extraction, preflight classifies known
terminal cases. Conditions using unsupported \code{pe.*} or \code{math.*}
module semantics are \emph{unsupported}. A prescribed whole-file cryptographic
digest is \emph{infeasible} because construction would require a hash preimage.
A value that cannot fit its integer width is \emph{unsatisfiable}.
Preflight also checks the i686 MinGW compiler for the narrow supported PE32
predicate; other missing compilers surface later as construction errors. These
classes prevent a model fallback from guessing around a known semantic or
physical impossibility.

\subsection{String Evidence}

The extractor first parses declarations and then scans the condition token
stream to identify required evidence. Direct positive references require one
occurrence; immediate \code{not \$id} references are excluded. Simple
\code{\#id == n}, \code{\#id > n}, and \code{\#id >= n} conditions produce an
occurrence lower bound. The equality case places $n$ intended witnesses but does
not prove that headers, other strings, or padding create no additional match.
For \code{all of them} or a supported wildcard set, every eligible declaration
is required. For \code{any}, the first declaration in
stable order is selected unless another member is already required. This is a
deterministic sufficient witness for the existential condition.

Literal \code{at} constraints are collected per identifier; multiple offsets
produce multiple typed entries. One literal \code{in (l..u)} range per
identifier is retained for the allocator. Exact placements have priority over
ranged and unconstrained requests. A conflict in which an identifier is both
required and explicitly negative is reported as unsatisfiable rather than
delegated to a model. These decisions are token-based: only an immediately
preceding \code{not} is recognized, so more complex negation is outside the
sound fragment.

Fixed text is decoded from its YARA source representation. Ordinary strings use
UTF-8, wide-only ASCII strings use a UTF-16LE-compatible representation, and
fixed hex declarations preserve their exact bytes. When control or undecodable
bytes prevent a faithful text representation, the extractor retains the
declaration as hex. An \code{ascii wide} declaration permits the lower-cost
ASCII witness.

Wide-only non-ASCII and control-byte literals are rejected by deterministic
extraction. Non-fixed regular expressions and hex patterns instead require
normalization or extraction fallback when recognized by the capability
classifier. The classifier is incomplete, so this mechanism does not detect
every unsupported YARA form.

\subsection{Integer Evidence}

The deterministic parser recognizes a fixed-width integer call followed by a
literal comparison. It turns a predicate into one concrete witness. For a
non-negative compared value $v$, the selected value is
\begin{equation}
W(op,v)=
\begin{cases}
v       & \text{if } op\in\{=,\geq,\leq\},\\
v+1     & \text{if } op \text{ is } {>},\\
v-1     & \text{if } op \text{ is } {<} \text{ and } v>0,\\
0       & \text{if } op \text{ is } \neq \text{ and } v\neq 0,\\
1       & \text{if } op \text{ is } \neq \text{ and } v=0.
\end{cases}
\end{equation}
The result must fit the outer read width. Unsupported arithmetic and predicates
without a non-negative constructible witness terminate deterministic extraction.
Big-endian functions retain their byte order for later serialization.

The sound nested subset uses an inner little-endian \code{uint32} pointer and
one outer read. The parser currently accepts a broader family but does not retain
the inner width or byte order, so those broader forms are excluded from the
claimed subset. Listing~\ref{lst:nested-yara-read} illustrates the supported
form by reading the PE optional-header magic relative to the pointer stored at
file offset \code{0x3c}.
\begin{minipage}{\linewidth}
\begin{lstlisting}[language=YARA,caption={Nested YARA read represented by the integer extractor.},label={lst:nested-yara-read}]
rule pe32_optional_header {
  condition:
    uint16(uint32(0x3c) + 0x18) == 0x010b
}
\end{lstlisting}
\end{minipage}
The predicate becomes an entry with base offset \code{0x3c}, outer width two,
nested flag set,
and relative offset \code{0x18}. When PE routing is independently established,
the PE backend reads the actual pointer in the generated image and applies the
witness relative to that target. Other backends can instead construct a
synthetic file-level pointer relation.

\subsection{Extraction Fallback}

Normalization and extraction fallback are separate boundaries. If a
deterministic extractor explicitly rejects unsupported syntax, the extraction
role may return the same typed schemas used by the deterministic path. Structured
output is attempted first; prompt-injected JSON is used when the provider lacks
tool-call support. Schema validation constrains shape, not semantic completeness,
so a fallback can still omit evidence or choose an incorrect offset.

Fallback cannot override a successful deterministic parse. Model-selected values
can influence serialized payload bytes, but conventional code retains exclusive
control over source templates, container structure, placement, and serialization.
This separation keeps both normalization choices and fallback evidence on the
data side of the trust boundary.

\subsection{Layout Allocation}

Where the shared allocator is used (ELF, generic, and scanner-oriented paths),
every string entry is serialized before placement. The allocator expands
occurrence counts and orders requests by exact, ranged, then unconstrained
placement. Runnable PE has separate placement logic and does not realize every
range or count represented by the shared schema. For a witness interval
$[o,o+L)$ and a \code{fullword} boundary width $b$, availability is tested over
\begin{equation}
  [\max(0,o-b), o+L+b).
\end{equation}
Exact collisions fail. Ranged placements advance one byte until a free position
is found. Unconstrained witnesses are packed from a backend-specific base with a
one-byte gap. Direct generic construction additionally reserves constant ranges
before allocating strings.

This allocator is deterministic for fixed typed input. Complete artifact bytes
may still vary because compiler/linker versions influence generated containers
and filesize padding uses random bytes.

\section{Deterministic Binary Realization}
\label{sec:realization}

\begin{figure*}[t]
  \centering
  \includegraphics[width=0.95\textwidth]{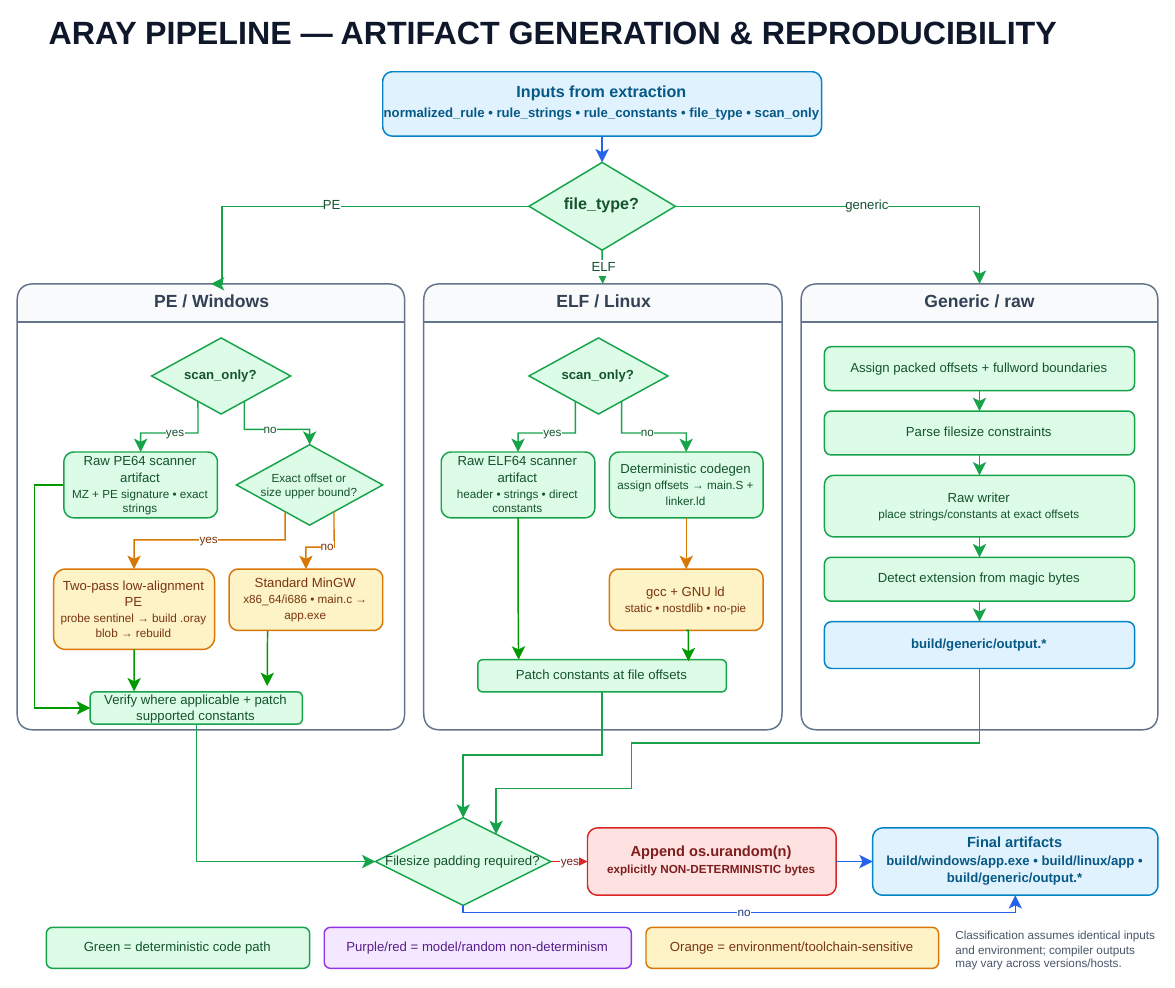}
  \caption{Artifact realization paths. Runnable ELF and PE outputs are generated
  by deterministic code but remain toolchain-sensitive. Scanner-oriented writers
  bypass the compilers. Random filesize padding is the only intentional source
  of byte-level randomness. End-to-end acceptance is established separately by
  scanning the final artifact with YARA against the independently selected
  upstream original rule.}
  \label{fig:backends}
\end{figure*}

Figure~\ref{fig:backends} summarizes the construction paths. Routing uses both
typed entries and selected lexical cross-checks against the normalized rule.
Native MZ or PE-signature evidence triggers the PE route; ordinary wide strings
act as an additional PE heuristic. Fixed non-PE magic at offset zero, or a low
flat placement, triggers the generic writer. ELF is the default route.

These decisions concern raw bytes and typed predicates. They do not implement
the semantics of an imported YARA module such as \code{pe}. Generic offset-zero
evidence is cross-checked against source text, whereas PE-signature nesting and
wide-format claims do not receive an equally strong check. Wide data is not
intrinsically Windows-specific.

\subsection{Binary-Format Terminology}

A \emph{file offset} is a zero-based byte index in the serialized artifact. A
virtual address (VA) denotes a location after loading; a linker's virtual memory
address (VMA) is the address assigned while laying out the image. In PE, a
relative virtual address (RVA) is measured from the image base. Alignment rounds
file or memory regions to boundaries required by the format or loader. These
coordinate systems need not coincide, which is why YARA's file-oriented
\code{at} operator cannot be implemented by virtual placement alone.

Table~\ref{tab:binary-structures} summarizes the structures used by the
backends. In ELF, sections organize link-time content, while segments describe
the regions consumed by the loader. In PE, the section table carries both an
RVA and a raw file location. The so-called Optional Header is optional for COFF
object files but required for executable PE images \cite{sysvabi,microsoftpecoff}.

\begin{table*}[t]
  \small
  \centering
  \begin{tabular}{@{}>{\raggedright\arraybackslash}p{0.18\textwidth}>{\raggedright\arraybackslash}p{0.34\textwidth}>{\raggedright\arraybackslash}p{0.40\textwidth}@{}}
    \toprule
    \textbf{Structure} & \textbf{Relevant fields} & \textbf{Role in Aray} \\
    \midrule
    ELF header & Class, byte order, machine, entry point, and table offsets & Identifies ELF64 and locates loader metadata; the direct writer emits only this 64-byte header. \\
    ELF program header & \code{p\_offset}, \code{p\_vaddr}, sizes, flags, and alignment & A \code{PT\_LOAD} segment defines the mapping from file bytes to virtual memory. \\
    ELF section header & \code{sh\_offset}, \code{sh\_addr}, type, and flags & Describes link-time sections; runnable output uses named sections, but the loader follows segments. \\
    PE DOS header & \code{MZ} and the \code{e\_lfanew} file pointer & Establishes the DOS-compatible prefix and locates the PE signature. \\
    PE signature and COFF header & \code{PE\textbackslash0\textbackslash0}, machine, and section count & Identifies the image and records its target machine and section count. \\
    PE Optional Header & Magic, entry point, image base, alignments, and data directories & Defines image-wide loader properties and the PE32/PE32+ variant. \\
    PE section table & RVA, virtual size, \code{PointerToRawData}, and raw size & Relates section bytes on disk to their location in the loaded image. \\
    \bottomrule
  \end{tabular}
  \caption{ELF and PE structures relevant to file-layout synthesis.}
  \label{tab:binary-structures}
\end{table*}
\subsection{Runnable ELF: Assembly, Sections, and Linker Contract}

The Linux backend avoids C and libc. For every placed byte range $(F_i,B_i)$,
the generator emits a dedicated GNU assembler section, as shown in
Listing~\ref{lst:elf-asm-section}.
\begin{lstlisting}[language=GNUAssembler,caption={Generated ELF payload section for a witness requested at offset \code{0x600}.},label={lst:elf-asm-section}]
# Bytes required by the selected YARA rule.
.section .sec_0x600,"aw",@progbits
.byte 0x61, 0x6c, 0x70, 0x68, 0x61
\end{lstlisting}
It also emits an \code{\_start} routine that invokes the x86-64 Linux
\code{write} and \code{exit} system calls. The resulting source is linked with
\code{gcc -static -nostdlib -no-pie}.

Ordinary section placement controls virtual addresses, whereas YARA's
\code{at} operator refers to file offsets. \tool bridges these coordinate
systems with the explicit program-header contract in
Listing~\ref{lst:elf-linker-script}.
\begin{lstlisting}[language=LinkerScript,style=araynumbered,caption={Generated GNU linker script for two exact-offset sections.},label={lst:elf-linker-script}]
PHDRS { load PT_LOAD FILEHDR PHDRS ; }
SECTIONS
{
    . = 0x400000 + SIZEOF_HEADERS;
    . = 0x400600;
    .sec_0x600 : { *(.sec_0x600) } :load
    . = 0x400800;
    .sec_0x800 : { *(.sec_0x800) } :load
    .text : { *(.text*) } :load
}
\end{lstlisting}
For an ELF load segment, a byte at file offset $F$ is mapped according to
\begin{equation}
  \mathrm{VA}=p_{\mathrm{vaddr}}+(F-p_{\mathrm{offset}}).
\end{equation}
Declaring one \code{PT\_LOAD} with \code{FILEHDR PHDRS} causes the file header
and program headers to belong to the segment, establishing
$p_{\mathrm{offset}}=0$ and $p_{\mathrm{vaddr}}=0x400000$. Placing a section at
$\mathrm{VMA}=0x400000+F$ therefore yields
\begin{equation}
  F=\mathrm{VMA}-0x400000.
\end{equation}
The explicit program header is essential: relying on a linker's default script
may introduce a page-sized initial file offset and invalidate every YARA
placement constraint \cite{sysvabi,gnuld}. Generated \code{main.S} and
\code{linker.ld} expose this contract for inspection.

After linking, direct integer witnesses are patched at their file offsets using
the extracted width and byte order. For a nested file-level witness, the backend
can append a target value and write its file offset as a 32-bit little-endian
pointer at the inner location. This satisfies YARA's file dereference but does
not imply that the appended \emph{overlay}---bytes beyond the mapped
segments---is available for runtime access.

\subsection{Runnable PE: Native Structure and Ordinary Strings}

For PE rules without exact string offsets or a filesize upper bound, \tool emits
deterministic C globals and invokes MinGW. Fixed hex data uses byte arrays.
Ordinary text uses string-initialized byte arrays, while wide text uses
\code{wchar\_t} globals. A normal \code{main} prints a fixed banner and exits.

The compiler-generated image natively provides \code{MZ} at offset zero, an
\code{e\_lfanew} pointer at \code{0x3c}, and \code{PE\textbackslash0\textbackslash0}
at the referenced location. Here, \code{e\_lfanew} is the DOS-header field that
stores the PE signature's file offset. This native layout satisfies the common
predicate in Listing~\ref{lst:pe-header-predicate}.
\begin{lstlisting}[language=YARA,caption={YARA predicate for the native MZ and PE signatures.},label={lst:pe-header-predicate}]
rule native_pe_header {
  condition:
    uint16(0) == 0x5a4d and
    uint32(uint32(0x3c)) == 0x00004550
}
\end{lstlisting}
PE32 uses a 32-bit optional header and the i686 compiler; PE32+ uses the 64-bit
header and the x86-64 compiler. Selected computed predicates determine which
variant is required. Nested relative constants are patched by following the
pointer in the generated file rather than assuming a fixed PE-header location.

The ordinary PE path does not patch direct non-structural constants. Such rules
require the offset-aware or direct-writer strategy. The router also treats
PE-signature bytes extracted at \code{0x3c} as PE evidence without checking the
nested flag, which can misroute an unusual direct predicate.

\subsection{Offset-Aware PE: Low Alignment and Two Passes}

A conventional PE link does not guarantee that a C global will appear at an
arbitrary file offset. The offset-aware backend therefore emits a freestanding C
program with a dedicated \code{.oray} section. Listing~\ref{lst:pe-link-command}
shows the complete low-alignment link command for PE32+.
\begin{lstlisting}[language=bash,caption={MinGW command for the offset-aware PE32+ backend.},label={lst:pe-link-command}]
x86_64-w64-mingw32-gcc \
  -o app.exe main.c \
  -nostdlib -lkernel32 -e _start \
  -Wl,--image-base,0x140000000 \
  -Wl,--file-alignment,0x10 \
  -Wl,--section-alignment,0x10
\end{lstlisting}
The custom \code{\_start} calls \code{GetStdHandle}, \code{WriteFile}, and
\code{ExitProcess}; no C runtime is required. With equal low file and section
alignment, MinGW produces a flat relation in which a section's raw file pointer
and RVA coincide. For raw pointer $P$, RVA $R$, and byte index $j$,
\begin{equation}
  F=P+j,\qquad \mathrm{RVA}_{byte}=R+j,
\end{equation}
and the intended invariant $P=R$ gives $F=\mathrm{RVA}_{byte}$.

Because the linker's section prefix is not predicted, construction uses two
passes. The first build places an eight-byte sentinel in \code{.oray}; searching
the probe binary yields its file offset $D$. A sentinel is a distinctive marker
used only to discover placement; Listing~\ref{lst:oray-section} shows its source
representation. For each requested witness
$(F_i,B_i)$ with $F_i\ge D$, the second-pass blob is defined by
\begin{minipage}{\linewidth}
\begin{lstlisting}[language=C,caption={Generated C declaration that places the probe or final blob in \code{.oray}.},label={lst:oray-section}]
__attribute__((section(".oray"), used))
unsigned char _oray[] = {
  0xAB, 0x61, 0x72, 0x61, 0x79, 0x53, 0xCD, 0xEF
};
\end{lstlisting}
\end{minipage}
\begin{equation}
  \mathrm{blob}[F_i-D:F_i-D+|B_i|]=B_i,
\end{equation}
with zero-filled gaps. Requests below $D$ are accepted only if the probe's native
headers already contain the required bytes. The program is rebuilt with the
final blob, and every exact string interval is read back and compared byte for
byte. This verification checks YARA-visible file placement; it does not by
itself prove loader behavior or protect against a subsequent overlapping
constant patch. MinGW output can also contain build timestamps, so identical
inputs and toolchain versions do not imply byte-identical PE files.

\subsection{Scanner-Oriented and Generic Writers}

The \code{--scan-only} path bypasses GCC and MinGW. Its ELF writer emits a
64-byte ELF64 header followed by direct byte ranges. Because it emits neither
program nor section headers, the result is not runnable. This path also omits
nested constants.

The scan-only PE writer emits the native identification chain described in
Table~\ref{tab:binary-structures}, then places string witnesses directly. It
does not currently receive extracted constants. Moreover, an exact placement at
a low offset can overwrite the DOS header, PE signature, COFF header, or
Optional Header. The result is intended for raw scanning, not execution or
general loader compatibility.

Non-PE magic and low flat layouts use the generic writer. It reserves integer
ranges in a zeroed buffer before packing unconstrained strings from offset
\code{0x10}. Constants and synthetic nested pointers are then written directly.
Known magic bytes determine the output extension, but do not establish that the
blob is a valid document or archive. Direct writers use a minimum size of
\code{0x200}, so they cannot satisfy smaller filesize upper bounds.

\subsection{Filesize Constraints and Acceptance Oracle}

Textual filesize comparisons are reduced to a half-open interval $[m,M)$.
If the initial artifact size $s$ is below $m$, a target inside the interval is
chosen and bytes are appended. The amount of padding is a deterministic
function of $(s,m,M)$, but the contents use \code{os.urandom}; padded artifacts
are therefore not byte-for-byte reproducible and random bytes can theoretically
interact with negative or count-sensitive predicates. If $s\ge M$, the backend
warns and cannot shrink the artifact. Parsing is currently regex-based over the
rule text and does not reject contradictory bounds, so this mechanism is not a
general filesize constraint solver.

Construction completion is not the acceptance criterion. The regular CLI emits
an artifact without scanning it; the evaluation driver invokes the external YARA
engine with an independently associated upstream source rule. Association
requires the source filename, trailing directory path, and first public rule
identifier to agree; missing, ambiguous, or name-mismatched associations fail
explicitly rather than falling back to the normalized derivative. The selected
source rule is reduced to its reachable standalone dependency closure, and the
scan is restricted to that identifier so that an unrelated rule cannot produce a
false pass. A result is recorded as passed only when this original source
predicate matches the generated artifact. This oracle independently tests the
concrete synthesized witness against $r$, while still not proving universal
implication between $r'$ and $r$. Its independence is at the predicate level:
the evaluator still shares Aray's source-selection and dependency-closure
machinery and uses the same YARA implementation.

\section{Evaluation}
\label{sec:evaluation}

\subsection{Staged End-to-End Design and Corpus}

We ask three sequential questions: how many selected source entries Aray's
normalization workflow admits, how many accepted outputs pass constructibility
preflight, and, among those admitted to construction, how often deterministic
extraction and realization produce an artifact accepted by the independently
selected upstream original rule? We report \emph{normalization admission}, the
overall \emph{positive-fixture yield}, and \emph{conditional realization},
respectively. The two stages are measured separately to attribute model-dependent
normalization and deterministic realization, but they form one lineage from
public source files to terminal Aray dispositions.

The source corpus consists of 416 files from the public Yara-Rules repository
\cite{yararules} at commit
\texttt{\seqsplit{0f93570194a80d2f2032869055808b0ddcdfb360}}. In stage one,
\code{aray-normalize} reduced each file to its selected standalone target and
applied Aray's normalization and validation workflow. With
\code{glm-5.2:cloud} configured for normalization and residual judging, 182
entries followed the deterministic no-normalization path and 234 received model
normalization. All 416 outputs were accepted, for 100\% normalization admission.
Here, \emph{accepted} means that Aray's deterministic checks and, where needed,
its configured judge admitted the candidate; it does not mean human review or a
formal proof of equivalence. Before the final freeze, an original-rule scan
exposed two anchored-regex counterexamples that a derivative-only oracle had
missed. We strengthened the contextual regex validator, repaired those two
specializations, and retained their boundary patterns as regression tests. All
416 final original--derivative pairs then returned no error under the supported
deterministic syntax and semantic checks. This does not mean that all 416
transformations belonged to the narrow implication fragment that bypasses
residual judging. Because this refinement used failures observed in the same
corpus, the final result is a post-fix systems evaluation rather than a held-out
estimate of normalization generalization.

Those exact outputs were frozen before stage two. The SHA-256 digest of their
sorted file-hash manifest is
\texttt{\seqsplit{2b7d09de6a0f152218a471ad018dd50fe9af4a4aae08d5ac7fccfa9cd2cb985d}}.
Freezing creates an auditable handoff and prevents model-dependent regeneration
from obscuring realization failures; it does not make normalization external to
Aray. The frozen outputs were evaluated with \code{scan\_only=false} and one
worker. The environment provided GCC, both x86-64 and i686 MinGW, and the YARA
CLI. Generic routes remained direct byte writers despite the full-compile
setting.

To make accidental inference in stage two observable, both model roles were
configured with the name \code{deterministic-control} and an OpenAI-compatible
endpoint at \code{127.0.0.1:9}, which was expected to be unreachable. Any model
invocation would therefore fail the run. A match required successful construction
followed by positive output from the YARA CLI against the associated upstream
original predicate. The evaluator ran with original-rule validation enabled and
the upstream repository as its association root. The final control report records
11.418 seconds as the sum of per-rule durations; this is not a
hardware-independent performance benchmark.

\subsection{End-to-End Case Study}

Listing~\ref{lst:case-study-rule} gives a compact case that exposes each stage
without requiring normalization. It is a focused repository example, separate
from the aggregate corpus measurement.
\begin{lstlisting}[language=YARA,caption={Two exact-offset strings used for the end-to-end case study.},label={lst:case-study-rule}]
rule multi_offset {
  strings:
    $a = "alpha"
    $b = "beta"
  condition:
    $a at 0x600 and $b at 0x800
}
\end{lstlisting}
Source reduction selects \code{multi\_offset}, and the capability classifier
takes the no-normalization path. Deterministic extraction produces two ASCII
entries with exact file offsets \code{0x600} and \code{0x800}; there are no
integer entries. With no PE or non-PE magic predicate, routing selects the ELF
backend. Code generation emits sections for the two byte ranges, and the linker
contract places their VMAs at \code{ELF\_BASE+0x600} and
\code{ELF\_BASE+0x800}, making the corresponding file offsets exact.

We reproduced the complete path with an unreachable model endpoint. The output
was a statically linked ELF64 executable. Reading five bytes at \code{0x600}
returned \code{61 6c 70 68 61} (\code{alpha}), and reading four bytes at
\code{0x800} returned \code{62 65 74 61} (\code{beta}). Scanning with YARA
reported \code{multi\_offset build/linux/app}; executing the fixture printed the
fixed \tool banner and exited. This case demonstrates the concrete chain from
rule predicate, through typed evidence and layout, to a runnable matching
artifact. It does not imply that the artifact is the only member, or a preferred
member, of $\mathcal{M}_r$.

\subsection{Results}

The upper panel of Table~\ref{tab:results} reports the Aray normalization stage;
the lower panel reports terminal dispositions after its frozen outputs entered
deterministic realization. Stage one admitted all 416 entries. In stage two,
preflight admitted 406, and all 406 produced matching artifacts. The staged
positive-fixture yield was therefore 406/416 (97.6\%), while conditional
realization was 406/406 (100\%). The other ten terminated during constructibility
preflight: seven used unsupported \code{pe.*} module semantics, two prescribed
whole-file hash values, and one required an integer value outside the declared
width. Every admitted artifact matched the independently associated upstream
original rule. There were zero \code{yara\_mismatch} outcomes and no construction
failures among admitted rules.

\begin{table}[t]
  \footnotesize
  \centering
  \begin{tabular}{@{}lrr@{}}
    \toprule
    \textbf{Stage and outcome} & \textbf{Rules} & \textbf{Share} \\
    \midrule
    \multicolumn{3}{@{}l}{\textit{Normalization: 416 accepted}} \\
    No model normalization & 182 & 43.75\% \\
    Model-normalized & 234 & 56.25\% \\
    \midrule
    \multicolumn{3}{@{}l}{\textit{Stage-two terminal disposition}} \\
    Upstream original matched & 406 & 97.6\% \\
    Expected: unsupported \code{pe.*} & 7 & 1.7\% \\
    Expected: whole-file hash & 2 & 0.5\% \\
    Expected: integer width & 1 & 0.2\% \\
    \code{yara\_mismatch} & 0 & 0.0\% \\
    Construction failure & 0 & 0.0\% \\
    \midrule
    Total & 416 & 100.0\% \\
    \bottomrule
  \end{tabular}
  \caption{Sequential stages of the Aray evaluation. Both panels follow the same
  416-entry lineage and are not additive. Normalization shares are exact to two
  decimal places; realization dispositions are rounded to one decimal place. A
  pass requires a match against the independently associated upstream original
  rule. Positive-fixture yield was 406/416 and conditional realization was
  406/406.}
  \label{tab:results}
\end{table}

Because normalization had already been performed by Aray in stage one, every
frozen input correctly bypassed it in stage two. Both deterministic extractors
processed each of the 406 rules that passed preflight; extraction fallback was
invoked zero times. The remaining ten did not reach extraction. Thus all 416
frozen inputs reached a terminal disposition without contacting the stage-two
control endpoint. Among matches, evaluator output paths identify 258 ELF
artifacts, 101 PE artifacts, and 47 generic blobs. The first two groups exercised
compiler/linker paths; generic output remained a direct writer. This experiment
did not exercise either \code{--scan-only} writer, and the retained report does
not separate ordinary from offset-aware PE.

The original-rule oracle materially affected validation practice. An earlier
run produced 404 matches and two \code{yara\_mismatch} outcomes. In
\code{MALW\_AlMashreq}, a doubly anchored wide regex had been replaced by a free
literal inside a seven-of-nine specialization. In \code{RAT\_PoetRATDoc}, the
shortest selected regex branch depended on end-of-file placement. The contextual
validator now rejects both proposals: the repaired AlMashreq specialization
selects seven non-anchored declarations, while PoetRAT selects the unanchored
\code{Python} alternative. Both repaired fixtures pass deterministic candidate
validation and the final upstream-original oracle. Their boundary patterns are
also covered by self-contained regression tests; the complete frozen corpus is
not versioned.

\subsection{Interpretation}

Taken together, the two stages measure workflow completion from Aray's
deterministic source selection and optional normalization through preflight,
extraction, realization, and final scanning. This includes toolchain-sensitive
ELF and PE realization as well as direct generic writing. The frozen boundary
also provides a component result: the deterministic core produced an artifact
accepted by the upstream original oracle for every constructible entry in this
corpus.

The end-to-end label is operational, not a claim of formally verified semantic
equivalence. The upstream-original scan can reveal concrete normalization or
realization errors that a derivative-only oracle would miss; it exposed the two
anchored-regex regressions above, and zero such mismatches remained in the final
run. This is empirical validation of one synthesized witness per constructible
source rule, not a solver-backed proof of universal subset implication or
coverage of arbitrary YARA.

The zero-mismatch result after preflight is methodologically important. It shows
that known unsupported or impossible cases can be separated from construction
errors instead of being reported as undifferentiated failures. The unreachable
endpoint is also stronger than inspecting configured model names: it turns an
unexpected invocation into an immediate experimental failure.

\section{Discussion}

\subsection{Determinism and Reproducibility}

For fixed source and typed input, interpretation, layout planning, and source
generation follow deterministic code paths. Routing and post-build patch plans
are deterministic as well. This is \emph{logical determinism}: the same
constraints lead to the same construction decisions.
Byte-for-byte reproducibility is narrower. Compiler and linker versions may
change container bytes and section placement, and random filesize padding is
explicitly non-reproducible. Exact-offset PE mitigates one toolchain uncertainty
by probing and verifying the final file instead of assuming a section origin.

\subsection{Safety and Intended Use}

Runnable-backend outputs are designed with minimal scaffolding that prints a
fixed banner and exits; this evaluation scanned but did not execute or
loader-validate every artifact. Scanner-oriented outputs are not executable.
Nevertheless, artifacts contain rule-selected bytes and may trigger security
controls by design. They should be labeled and handled as detection-test
fixtures. \tool reduces the need to distribute malware, but it does not make
every arbitrary YARA witness safe to execute or expose.

Operational use should therefore preserve a containment boundary even when the
fixture is designed to be non-malicious. A recovery exercise should use
dedicated buckets or prefixes, least-privilege and short-lived identities,
explicit replication scope, denied-by-default egress, centralized audit logs,
and an environment that can be rebuilt from a known state after the test. The
scanner need only read the fixture; no pipeline component should execute it.
Production credentials and unrelated datasets should remain outside the test
environment. These controls also bound optional model assistance: Aray exposes
no execution tools to the model, but it neither provisions nor validates the
surrounding sandbox, cloud account, or recovery topology.

\section{Limitations}

First, the lexical front end and extractor do not implement a complete YARA AST
or a general satisfiability solver. Complex Boolean or arithmetic forms can be
unsupported or over-constrained; loops and imported modules are also outside the
deterministic fragment. Rule-set reduction chooses one sufficient branch rather
than preserving every alternative. Because normalization and fallback triggers
are incomplete, parenthesized negation or an ignored modifier can still proceed
with incomplete evidence.

Second, witness synthesis is existential rather than one-to-one. A regex,
variable hex pattern, disjunction, or unconstrained placement can describe many
members of $\mathcal{M}_r$, potentially infinitely many. \tool returns one
candidate selected by stable-order, minimum-length, and feasibility heuristics;
it does not prove that this candidate is minimal, canonical, representative of
the detected family, or preferable to the alternatives. Conversely, different
rules can admit the same fixture. A successful match therefore certifies one
positive witness, not an exhaustive realization of the rule.

Third, model normalization is approximate. Deterministic checks establish
syntax and rule identity, then verify retained declarations and witness
membership, including representative boundary contexts for retained regex
witnesses. General subset implication remains judged rather than proven, and an
\code{uncertain} verdict currently proceeds in the regular graph; the batch
normalizer used by the evaluation retains only \code{passed} candidates. The
upstream-original oracle independently validates the concrete artifact generated
from each constructible candidate, but does not turn that observation into a
universal implication proof. The oracle is independent of the normalized
construction predicate, but shares Aray's source-selection and dependency-closure
machinery and the same YARA implementation; its independence is therefore
predicate-level, not implementation-level. Extraction fallback validates its
schema but has no complete semantic oracle before construction.

Fourth, interval integrity is incomplete across all backends. String placement is
collision-checked, but a later constant patch can overlap a string or structural
byte. PE read-back occurs before that patch, and constant--constant conflicts are
not globally solved. Separately, exact occurrence equality is represented only
as an intended lower bound, so incidental bytes can add matches. Integer support
also remains incomplete for signed \code{int16} reads and nested inner functions
other than little-endian \code{uint32}.

Fifth, backend support is uneven. Ordinary runnable PE relies on compiler
retention for free globals and does not realize every range or occurrence count;
it also skips direct non-structural constants. Scan-only PE forwards no extracted
constants, always emits PE32+, and permits low strings to overwrite the header
structures listed in Table~\ref{tab:binary-structures}. Scan-only ELF omits
nested constants. Generic output is a byte witness rather than a structurally
valid document. None of the scanner-oriented outputs should be treated as a
runnable format implementation.

Sixth, filesize padding can only enlarge artifacts and uses random data. Tight
upper bounds can be impossible for a selected full-compile backend, and random
padding can alter count or absence predicates. Textual parsing does not reject
contradictory bounds. Toolchain versions and PE timestamps also limit
byte-for-byte reproducibility.

Finally, the 406/416 result is the positive-fixture yield of a staged Aray
workflow whose normalization phase used GLM-5.2 and whose realization phase used
a frozen handoff. It is end to end under an upstream-original-rule oracle: 406
artifacts matched, ten cases ended in expected preflight dispositions, and no
constructible case produced \code{yara\_mismatch}. The two regex regressions used
to strengthen the validator were discovered in this corpus before the final run,
so the result is not a held-out estimate. Nor should it be interpreted as formal
normalization fidelity, full coverage of unmodified public YARA repositories, or
a language-model benchmark. The repository currently versions aggregate
provenance and a corpus digest, not the frozen files or canonical per-rule
reports, which limits independent reproduction from a clean clone.

\section{Related Work}

Research on YARA has primarily addressed scalable intelligence and detector
generation. YARIX organizes rule-based malware intelligence at scale
\cite{brengel2021yarix}, but does not construct positive objects for testing an
ingestion path. AutoYARA infers rules from samples through biclustering
\cite{raff2020autoyara}; PackGenome generates packer signatures
\cite{li2023packgenome}; NeuroYara ranks generated rules
\cite{mansour2025neuroyara}; and APIARY derives rules from behavioral API data
\cite{coscia2025apiary}. Living off the Analyst extracts rule features for a
separate classifier \cite{gupta2024living}. Their principal direction is from
samples, behavior, or existing rules toward a detector. \tool reverses that
direction: it treats an existing rule as the specification, constructs one
positive file-level witness, and preserves the construction predicate for
inspection. In the reported evaluation, the evaluator additionally uses the
derivative-independent upstream original rule as its acceptance oracle.

Recent work has used language models to extract security knowledge or generate
detection rules from reports and packages
\cite{schwartz2025llmcloudhunter,hu2024idsagent,zhang2025rulellm,buechel2025sok}.
Those systems place learned generation on the detector-construction path. \tool
instead uses a model only behind an explicit normalization/fallback boundary and
keeps supported extraction, source templates, layout, and serialization under
conventional code. Binary realization is governed by exact offsets, integer
widths, alignments, and container invariants; encoding these operations as code
makes them testable and permits read-back verification. Model-selected literals
remain untrusted payload input. This follows the program-synthesis principle of
separating a specification from a constrained realizer
\cite{gulwani2017programsynthesis}, while avoiding unconstrained generation at
the byte boundary.

LLM-assisted cyber evaluation presents a related but distinct containment
problem. ExploitGym packages realistic userspace, browser, and kernel targets in
controlled environments and reports that Claude Mythos Preview produced working
exploits for 157 of 898 instances; enabling standard mitigations reduced, but did
not eliminate, model success \cite{wang2026exploitgym}. Anthropic's agentic
misalignment study provides a different class of evidence: models with tool
access leaked information or took other harmful actions in deliberately
constrained simulations, while the authors explicitly report no corresponding
behavior observed in real deployments \cite{anthropic2025misalignment}. Kimi K3
has no comparable public containment incident; its primary documentation instead
reports long-horizon terminal orchestration and warns that excessive
proactiveness can produce unexpected decisions on a user's behalf
\cite{moonshot2026kimik3}. These results are therefore not interchangeable with
the observed OpenAI--Hugging Face intrusion. Together, however, they motivate
least privilege, independent action mediation, restricted egress, ephemeral
credentials, monitoring, and rebuildable environments whenever an LLM is given
tools in a security or recovery exercise. Aray adopts the narrower architectural
response of treating model output as untrusted declarative input and reserving
construction authority for deterministic code.

The work is also related to constrained input generation and program synthesis.
G2Fuzz synthesizes generators for non-textual fuzzing inputs
\cite{zhang2025g2fuzz}, whereas \tool does not seek broad input diversity or
fuzzing coverage. It realizes one existential witness for a selected YARA
predicate, including exact file offsets and executable-container constraints,
while the reported evaluation checks the result with the external YARA engine
against the associated upstream original predicate. Thus its novelty lies in
turning an operational detector into an auditable, format-aware positive fixture,
not in general-purpose file generation.

\section{Future Work}

The highest-priority improvement is a grammar-backed condition AST coupled to a
joint constraint solver. The solver should reason about content, placement,
multiplicity, absence, and filesize without assigning each evidence class in
isolation. A unified interval plan can then reserve strings and constants before
construction and verify them again after patching and padding. Deterministic
witness generation and contextual validation should also cover more regex and
nonlinear hex forms, while a formal or solver-backed subset check further narrows
the model boundary. Until that implication check is available, a strict mode
should stop normalization on an \code{uncertain} verdict rather than admitting
the candidate to construction.

The medium-term priority is module-aware synthesis. Although \tool generates PE
containers, it does not currently interpret conditions from \code{import "pe"}.
Initial support should translate selected structural fields, section properties,
and import/export predicates into typed PE constraints. A field-level capability
matrix should distinguish predicates that are constructible, merely verifiable,
infeasible, or not yet supported. The same architecture can later target
file-derived modules such as \code{elf}, \code{dotnet}, and \code{lnk}. Modules
that depend on external observations require a separate metadata contract, while
functions involving prescribed hashes must retain explicit infeasibility checks.

Backend verification should parse final ELF program headers and PE section tables
rather than rely only on construction-time assumptions. It should also forward
supported constants to scan-only PE and produce structurally valid generic
containers where format validity matters. Versioned toolchain environments and a
deterministic, predicate-aware padding mode would improve reproducibility.

In the longer term, \tool should support YARA-X, which is intended as YARA's
successor but is not API-compatible with YARA 4.x \cite{yaraxdocs}. A scanner
abstraction should accommodate both engines without assuming that rule-level
compatibility implies identical semantics. Differential tests should cover
syntax differences, module outputs, and match results. Reporting acceptance
against both engines would make compatibility an observed property rather than
an assumption.

Finally, future evaluations should retain the upstream-original oracle while also
reporting a derivative-rule scan. The pair helps attribute failures: a
derivative-only match is evidence of a normalization-fidelity problem, while
failure of the derivative indicates an extraction or realization problem. The
four possible outcomes should be reported explicitly rather than treated as a
complete binary diagnosis. Complete per-rule reports, source-rule associations,
artifact hashes, tool versions, and normalization-validation provenance should be
published. Replication across held-out repositories, commits, and YARA engines
would provide a stronger estimate of external validity than the post-fix corpus
reported here. Table-generation scripts should derive every aggregate result,
and boundary failures should remain separately attributable.

\section{Conclusion}

\tool reframes positive YARA validation as selecting and synthesizing one
file-level witness from a rule's match set.
The current system is best understood as a deterministic interpreter and binary
realizer with optional model-assisted normalization, not as an LLM binary
generator. Conventional code derives the supported string and integer evidence
and assigns its file layout. The ELF backend enforces an explicit mapping between
file offsets and virtual addresses. The offset-aware PE backend instead discovers
linker placement in a probe build and verifies requested bytes after rebuilding.

In the first stage of the evaluation, Aray's own normalization workflow accepted
all 416 selected public-rule entries: 182 (43.75\%) required no model
normalization, while 234 (56.25\%) were model-normalized. All final
original--derivative pairs returned no error under the supported deterministic
checks before the frozen handoff; this is distinct from proving every pair in the
narrow implication fragment. Preflight admitted 406, and all 406 artifacts matched
their independently associated upstream original rules, yielding 406/416
(97.6\%) end-to-end positive-fixture yield, 406/406 (100\%) conditional
realization, and zero \code{yara\_mismatch} outcomes. The other ten received
expected terminal dispositions, and no model endpoint was contacted during the
second stage. This independently validates the concrete witnesses without
claiming universal implication or general coverage of YARA. The results
demonstrate that a substantial and auditable deterministic core can turn
normalized YARA predicates into portable test fixtures without distributing the
corresponding malware. Such fixtures can
exercise object-ingestion and recovery paths without maintaining a matching
live-malware corpus in every test environment, reducing---but not eliminating---
the need for isolated, monitored, and resettable infrastructure.

\section{Availability}
\label{sec:availability}

\tool is open source under the Apache License 2.0 at
\begin{center}
  \url{https://github.com/c2dc/aray}
\end{center}
The reported evaluation used \tool
\href{https://github.com/c2dc/aray/releases/tag/v0.10.0}{\code{v0.10.0}} at commit
\href{https://github.com/c2dc/aray/commit/21682f53236975583ed38337d136d25b3f011cb7}{\texttt{\seqsplit{21682f53236975583ed38337d136d25b3f011cb7}}}.
The repository includes the interpretation pipeline and all binary backends,
together with their CLI and tests. It also versions diagram sources and a
machine-readable summary of both stages of the 416-entry evaluation, including
the original-rule oracle configuration and aggregate dispositions. The summary
is stored under \code{docs/site/assets/} as
\code{yara-rules-416-deterministic.json}.

\begingroup
\footnotesize
\bibliographystyle{plainnat}
\bibliography{references}

@inproceedings{brengel2021yarix,
  author    = {Michael Brengel and Christian Rossow},
  title     = {{YARIX}: Scalable {YARA}-based Malware Intelligence},
  booktitle = {30th USENIX Security Symposium (USENIX Security 21)},
  pages     = {3541--3558},
  year      = {2021}
}

@inproceedings{raff2020autoyara,
  author    = {Edward Raff and Richard Zak and Gary Lopez Munoz and William Fleming and Hyrum S. Anderson and Bobby Filar and Charles Nicholas and James Holt},
  title     = {Automatic {YARA} Rule Generation Using Biclustering},
  booktitle = {Proceedings of the 13th ACM Workshop on Artificial Intelligence and Security (AISec)},
  pages     = {71--82},
  year      = {2020},
  doi       = {10.1145/3411508.3421372}
}

@inproceedings{li2023packgenome,
  author    = {Shijia Li and Jiang Ming and Pengda Qiu and Qiyuan Chen and Lanqing Liu and Huaifeng Bao and Qiang Wang and Chunfu Jia},
  title     = {PackGenome: Automatically Generating Robust {YARA} Rules for Accurate Malware Packer Detection},
  booktitle = {Proceedings of the 2023 ACM SIGSAC Conference on Computer and Communications Security},
  pages     = {3078--3092},
  year      = {2023},
  doi       = {10.1145/3576915.3616625}
}

@article{mansour2025neuroyara,
  author  = {Ziad Mansour and Weihan Ou and Steven H. H. Ding and Mohammad Zulkernine and Philippe Charland},
  title   = {NeuroYara: Learning to Rank for {YARA} Rules Generation Through Deep Language Modeling and Discriminative N-Gram Encoding},
  journal = {IEEE Transactions on Dependable and Secure Computing},
  volume  = {22},
  number  = {2},
  pages   = {1747--1762},
  year    = {2025},
  doi     = {10.1109/TDSC.2024.3449641}
}

@article{coscia2025apiary,
  author  = {Antonio Coscia and Roberto Lorusso and Antonio Maci and Giuseppe Urbano},
  title   = {{APIARY}: An API-based Automatic Rule Generator for {YARA} to Enhance Malware Detection},
  journal = {Computers \& Security},
  volume  = {153},
  pages   = {104397},
  year    = {2025},
  doi     = {10.1016/j.cose.2025.104397}
}

@article{gupta2024living,
  author  = {Siddhant Gupta and Fred Lu and Andrew Barlow and Edward Raff and Francis Ferraro and Cynthia Matuszek and Charles Nicholas and James Holt},
  title   = {Living off the Analyst: Harvesting Features from {YARA} Rules for Malware Detection},
  journal = {arXiv preprint arXiv:2411.18516},
  year    = {2024},
  doi     = {10.48550/arXiv.2411.18516}
}

@inproceedings{schwartz2025llmcloudhunter,
  author    = {Yuval Schwartz and Lavi Ben-Shimol and Dudu Mimran and Yuval Elovici and Asaf Shabtai},
  title     = {{LLMCloudHunter}: Harnessing {LLM}s for Automated Extraction of Detection Rules from Cloud-Based {CTI}},
  booktitle = {Proceedings of the ACM Web Conference 2025},
  pages     = {1922--1941},
  year      = {2025},
  doi       = {10.1145/3696410.3714798}
}

@inproceedings{hu2024idsagent,
  author    = {Xiaowei Hu and Haoning Chen and Huaifeng Bao and Wen Wang and Feng Liu and Guoqiao Zhou and Peng Yin},
  title     = {A {LLM}-based Agent for the Automatic Generation and Generalization of {IDS} Rules},
  booktitle = {2024 IEEE 23rd International Conference on Trust, Security and Privacy in Computing and Communications (TrustCom)},
  pages     = {1875--1880},
  year      = {2024},
  doi       = {10.1109/TrustCom63139.2024.00259}
}

@inproceedings{zhang2025rulellm,
  author    = {XiangRui Zhang and XueJie Du and HaoYu Chen and Yongzhong He and Wenjia Niu and Qiang Li},
  title     = {Automatically Generating Rules of Malicious Software Packages via Large Language Model},
  booktitle = {2025 55th Annual IEEE/IFIP International Conference on Dependable Systems and Networks (DSN)},
  pages     = {734--747},
  year      = {2025},
  doi       = {10.1109/DSN64029.2025.00072}
}

@inproceedings{buechel2025sok,
  author    = {Marvin B{\"u}chel and Tommaso Paladini and Stefano Longari and Michele Carminati and Stefano Zanero and Hodaya Binyamini and Gal Engelberg and Dan Klein and Giancarlo Guizzardi and Marco Caselli and Andrea Continella and Maarten van Steen and Andreas Peter and Thijs van Ede},
  title     = {{SoK}: Automated {TTP} Extraction from {CTI} Reports -- Are We There Yet?},
  booktitle = {34th USENIX Security Symposium (USENIX Security 25)},
  pages     = {4621--4641},
  year      = {2025}
}

@inproceedings{zhang2025g2fuzz,
  author    = {Kunpeng Zhang and Zongjie Li and Daoyuan Wu and Shuai Wang and Xin Xia},
  title     = {Low-Cost and Comprehensive Non-textual Input Fuzzing with {LLM}-Synthesized Input Generators},
  booktitle = {34th USENIX Security Symposium (USENIX Security 25)},
  pages     = {6999--7018},
  year      = {2025}
}

@article{gulwani2017programsynthesis,
  author  = {Sumit Gulwani and Oleksandr Polozov and Rishabh Singh},
  title   = {Program Synthesis},
  journal = {Foundations and Trends in Programming Languages},
  volume  = {4},
  number  = {1--2},
  pages   = {1--119},
  year    = {2017},
  doi     = {10.1561/2500000010}
}

@techreport{bartock2016recovery,
  author      = {Michael Bartock and Jeffrey Cichonski and Murugiah Souppaya and Matthew Smith and Gregory Witte and Karen Scarfone},
  title       = {Guide for Cybersecurity Event Recovery},
  institution = {National Institute of Standards and Technology},
  number      = {NIST Special Publication 800-184},
  year        = {2016},
  doi         = {10.6028/NIST.SP.800-184}
}

@article{wang2026exploitgym,
  author  = {Zhun Wang and Nico Schiller and Hongwei Li and Srijiith Sesha Narayana and Milad Nasr and Nicholas Carlini and Xiangyu Qi and Eric Wallace and Elie Bursztein and Luca Invernizzi and Kurt Thomas and Yan Shoshitaishvili and Wenbo Guo and Jingxuan He and Thorsten Holz and Dawn Song},
  title   = {{ExploitGym}: Can {AI} Agents Turn Security Vulnerabilities into Real Attacks?},
  journal = {arXiv preprint arXiv:2605.11086},
  year    = {2026},
  doi     = {10.48550/arXiv.2605.11086}
}

@misc{openai2026hfincident,
  author       = {{OpenAI}},
  title        = {{OpenAI} and {Hugging Face} Partner to Address Security Incident During Model Evaluation},
  howpublished = {\href{https://openai.com/index/hugging-face-model-evaluation-security-incident/}{OpenAI incident report}},
  note         = {Published July 21, 2026; accessed August 16, 2026},
  year         = {2026}
}

@misc{huggingface2026timeline,
  author       = {{Hugging Face}},
  title        = {Anatomy of a Frontier Lab Agent Intrusion: A Technical Timeline of the July 2026 Incident},
  howpublished = {\href{https://huggingface.co/blog/agent-intrusion-technical-timeline}{Hugging Face technical timeline}},
  note         = {Published July 27, 2026; accessed August 16, 2026},
  year         = {2026}
}

@misc{anthropic2025misalignment,
  author       = {{Anthropic}},
  title        = {Agentic Misalignment: How {LLM}s Could Be Insider Threats},
  howpublished = {\href{https://www.anthropic.com/research/agentic-misalignment}{Anthropic research report}},
  note         = {Published June 20, 2025; accessed August 16, 2026},
  year         = {2025}
}

@misc{moonshot2026kimik3,
  author       = {{Moonshot AI}},
  title        = {{Kimi K3}: Open Frontier Intelligence},
  howpublished = {\href{https://www.kimi.com/blog/kimi-k3}{Kimi technical blog}},
  note         = {Published July 2026; accessed August 16, 2026},
  year         = {2026}
}

@misc{yaradocs,
  author       = {{VirusTotal}},
  title        = {{YARA} Documentation},
  howpublished = {\url{https://yara.readthedocs.io/}},
  note         = {Accessed August 15, 2026},
  year         = {2026}
}

@misc{yararules,
  author       = {{Yara-Rules Project}},
  title        = {Yara-Rules: A Collection of {YARA} Rules},
  howpublished = {\url{https://github.com/Yara-Rules/rules}},
  note         = {Corpus commit \texttt{\seqsplit{0f93570194a80d2f2032869055808b0ddcdfb360}}},
  year         = {2026}
}

@misc{sysvabi,
  author       = {{System V ABI Project}},
  title        = {System V Application Binary Interface: AMD64 Architecture Processor Supplement},
  howpublished = {\url{https://gitlab.com/x86-psABIs/x86-64-ABI}},
  note         = {Accessed August 15, 2026},
  year         = {2026}
}

@manual{gnuld,
  author       = {{Free Software Foundation}},
  title        = {{GNU ld}: The GNU Linker},
  organization = {GNU Project},
  howpublished = {\url{https://sourceware.org/binutils/docs/ld/}},
  note         = {Accessed August 15, 2026},
  year         = {2026}
}

@misc{microsoftpecoff,
  author       = {{Microsoft}},
  title        = {{PE} Format: Portable Executable and Common Object File Format Specification},
  howpublished = {\url{https://learn.microsoft.com/en-us/windows/win32/debug/pe-format}},
  note         = {Accessed August 15, 2026},
  year         = {2026}
}

@misc{yaraxdocs,
  author       = {{VirusTotal}},
  title        = {{YARA-X} Documentation},
  howpublished = {\url{https://virustotal.github.io/yara-x/}},
  note         = {Accessed August 16, 2026},
  year         = {2026}
}
\endgroup

\end{document}